\documentclass[twocolumn,showpacs,preprintnumbers,amsmath,amssymb,prl]{revtex4}

\usepackage[english]{babel}
\usepackage[latin1]{inputenc}
\usepackage{indentfirst}
\usepackage{graphicx}
\usepackage{amsmath}
\usepackage{amssymb}
\usepackage{amsthm}
\usepackage{mathrsfs}
\usepackage{lscape}
\usepackage{bm}
\usepackage{float}
\usepackage{longtable}
\usepackage{color}

\def \beq {\begin{equation}}
\def \eeq {\end{equation}}
\def \ba {\begin{eqnarray}}
\def \ea {\end{eqnarray}}

\begin{document}
\title{Phonon-induced dephasing of chromium colour centres in diamond}
\author{T. Müller$^{1}$, I. Aharonovich$^{2}$, Z. Wang$^{3}$, X. Yuan$^{3}$, S. Castelletto$^{4}$, S. Prawer$^{5}$, M. Atatüre$^{1}$}
\affiliation{$^{1}$Cavendish Laboratory, University of Cambridge, JJ Thomson Ave., Cambridge CB3 0HE, UK}
\affiliation{$^{2}$School of Engineering and Applied Science, Harvard University, Cambridge, MA 02138, USA}
\affiliation{$^{3}$HFNL\&Department of Modern Physics, University of Sci\&Tech of China, Hefei, 230026, China}
\affiliation{$^{4}$Centre for Quantum Science and Technology, Department of Physics and Astronomy, Macquarie University, Sydney, New South Wales 2109, Australia}
\affiliation{$^{5}$School of Physics, University of Melbourne, VIC 3010, Australia }

\vspace{-3.5cm}

\date{\today}
\begin{abstract}
We report on the coherence properties of single photons from chromium-based colour centres in diamond. We use field-correlation and spectral lineshape measurements to reveal the interplay between slow spectral wandering and fast dephasing mechanisms as a function of temperature. We show that the zero-phonon transition frequency and its linewidth follow a power-law dependence on temperature indicating that the dominant fast dephasing mechanisms for these centres are direct electron-phonon coupling and phonon-modulated Coulomb coupling to nearby impurities. Further, the observed reduction in the quantum yield for photon emission as a function of temperature is consistent with the opening of additional nonradiative channels through thermal activation to higher energy states predominantly and indicates a near-unity quantum efficiency at 4 K.
\end{abstract}
\pacs{61.72.J-, 81.05.ug} \maketitle

Diamond plays a key role in a wide range of applications in both electronics \cite{electronics}	 and photonics \cite{Greentree} today. Due to its wide bandgap it hosts a variety of optically active centres in the visible and the near infrared part of the spectrum \cite{Zaitsev}. The nitrogen-vacancy (NV) centre has attracted great attention due to its remarkable spin properties as a model system for quantum technologies \cite{Gruber, Wrachtrup, Degen}. Other optically active centres such as silicon-vacancy \cite{Neu}, nickel \cite{Nadolinny}, xenon \cite{Xenon} and chromium \cite{Igor1} have also shown desirable photonic properties such as short lifetimes, predominant emission into the zero-phonon line (ZPL), and spectral tuning via DC Stark effect \cite{Tamarat, Muller, Bassett}. However, the ZPL spectrum of diamond colour centres are typically broader than the radiative linewidth evidencing the influence of dephasing mechanisms on the optical transitions \cite{footnote1}. These dephasing mechanisms can range from slow spectral wandering of the transition due to charge fluctuations in the environment to fast dephasing processes. The former can be remedied by direct feedback control on the transition, which has been demonstrated for NV centres \cite{Acosta}. The latter is irreversible and  presents a fundamental limit to the single-photon coherence even with feedback. Therefore, it is essential to identify the source for these mechanisms, as well as the extent of their contributions to the spectral broadening of these centres. Here, we investigate the dephasing mechanisms influencing the chromium colour centres and report that the predominant sources for fast dephasing are direct electron-phonon coupling and phonon-modulated Coulomb coupling to nearby impurities, as manifested by the power-law temperature dependence of the transition frequency and linewidth. Further, the temperature dependence of the quantum yield reveals the existence of a thermal excitation mechanism to higher energy states similar to the neutral silicon-vacancy centres in diamond \cite{Warwick}.
\newline

\begin{figure}[t]
\centering
  \includegraphics[width=0.47\textwidth]{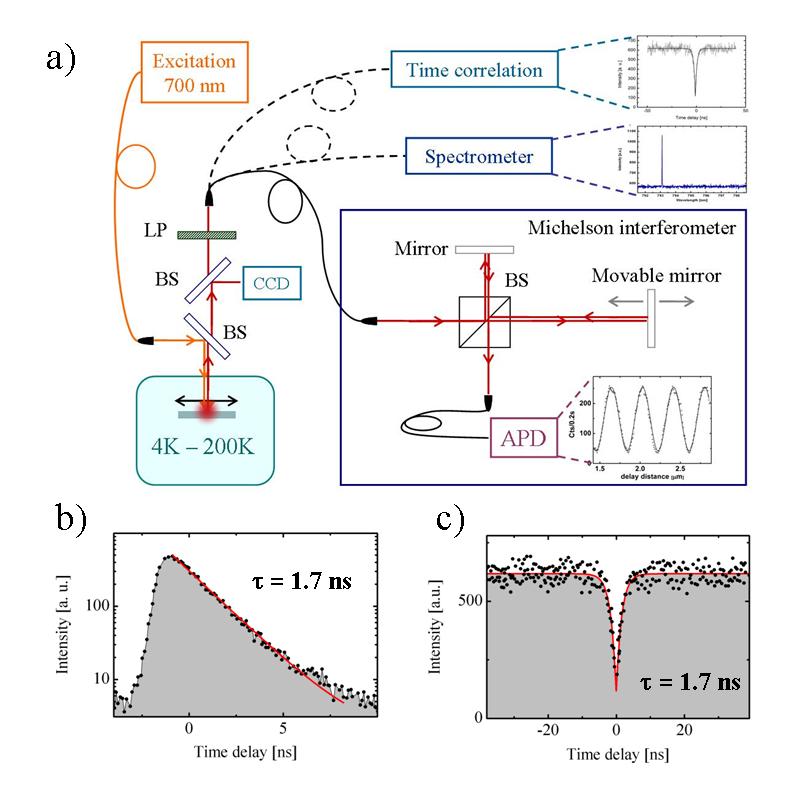}\\
  \caption{(Color online)(a) Schematic of the confocal microscope used for the experiments. BS is a beamsplitter and LP is a 740 nm longpass filter to remove residual laser reflection from luminescence by the chromium emitters. (b) Lifetime measured for the 773.3 nm transition line in figure 1a) inset. A value of 1.7 ns is extracted from the single exponential fit. (c) Autocorrelation measurement for the 773.3 nm line with g(2)(0) = 0.28. This value reflects the finite system response time and background emission from the crystal. The respective fit again reveals an excited state lifetime of 1.7 ns.
     }\label{fig1}
\end{figure}

Chromium centres can be formed either by chemical vapour deposition (CVD) growth, where the etched sapphire substrate acts as the source for chromium atoms \cite{Igor2}, or by implantation into bulk diamond \cite{Castelletto}. While the centres in bulk diamond exhibit narrower emission linewidth than those grown in nano- or microcrystals, they suffer from low photon collection efficiency due total internal reflection, as well as reduced photostability \cite{Muller}. We therefore concentrate on centres in microcrystals (size {$\sim$}2 microns) for the investigation of dephasing mechanisms. The experimental setup used for the optical measurements is illustrated in Fig. 1(a). A temperature-controlled cryostat and a homemade confocal optical microscope are used to access the centres optically. The chromium centres are excited by a continuous wave laser at 700 nm and the emission ($\sim 780$ nm) is collected into a single-mode fiber. A time-correlated photon counting setup is used for excited-state lifetime measurements, as well as intensity-correlation measurements. The excited-state lifetimes of chromium centres in bulk, micro- and nano-crystal diamonds can vary significantly \cite{Castelletto, Siyushev}. Figures 1(b) and 1(c) display exemplary results for these measurements performed at 4 K on a single chromium centre in a microcyrstal, and the value of 1.7 ns extracted from the measurements is consistent with the range of values reported to-date. Based on this value, transform-limited photon emission is expected to result in a coherence time of 3.4 ns. A calibrated and scanable Michelson interferometer and a spectrometer with 8-GHz resolution are used in parallel to quantify the degree of dephasing via the first-order coherence of the photon emission [see Fig. 1(a)].
\newline

\begin{figure}[t]
\centering
  \includegraphics[width=0.45\textwidth]{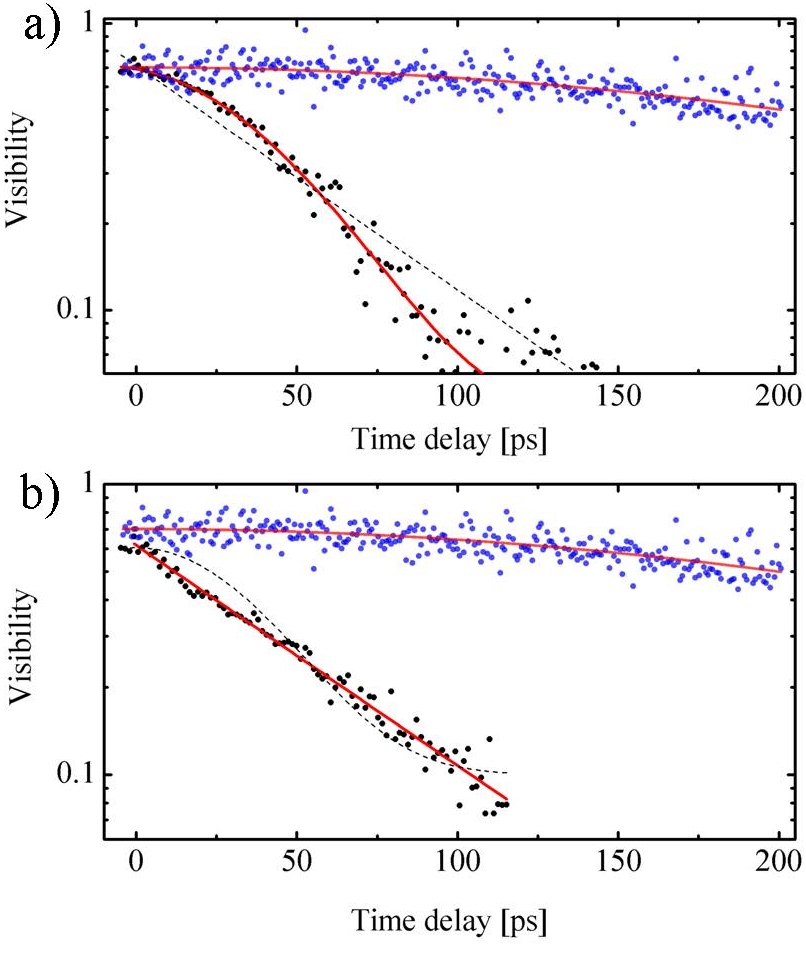}\\
  \caption{(Color online) First-order field autocorrelation measurement, $g^{(1)}$, performed at liquid helium temperature on centres A and B, as well as a chromium centre implanted in bulk diamond. (a) For centre A the visibility extracted from interference fringes at different time delays between the two arms of a Michelson interferometer (filled black circles) exhibits a Gaussian decay pattern with a $\tau_{1/e} = 62$ ps (solid red curve). An exponential function is shown for comparison (dotted black curve). The measured $g^{(1)}$ for a chromium centre located in bulk diamond (filled blue circles) display a coherence time longer than 700 ps (solid red curve). (b)  For centre B the extracted visibility (filled black circles) follows an exponential decay profile (solid red curve) with a coherence time of 57 ps. Again, a Gaussian profile (dotted black curve) is shown for comparison along with the $g^{(1)}$ of the chromium centre in bulk diamond. }
\label{fig:fig2}
\end{figure}

Figures 2(a) and 2(b) display the measured interference visibility $V = (I_{max}-I_{min})/(I_{max}+I_{min})$ as a function of relative time delay due to path-length difference for photons from two colour centres situated in two different microcrystals (labelled A and B, with ZPL at 756 nm and 770 nm, respectively). Filled black circles are the experimental data and the solid red curves are the theoretical fits. Common to both colour centres is  the coherence timescale  approximately 30 times lower than the predicted transform limit. The coherence functions for these centres are, however, fundamentally different from each other. Centre A shown in Fig. 2(a) displays a predominantly Gaussian decay of coherence (the solid red curve is a Gaussian fit with $\tau_{1/e} = 62$ ps), whereas for centre B shown in Fig. 2(b) the coherence decays exponentially (the solid red curve displays an exponential fit with $\tau_{1/e} = 57$ ps). The exponential form of the coherence function for centre B indicates a dominant contribution from the irreversible dephasing mechanisms to the emission spectrum, while the Gaussian-like coherence function of centre A limits an exponential contribution to a timescale not shorter than 210 ps. This corresponds to 750-MHz upper bound to the irreversible broadening for the linewidth for this centre. The relatively short coherence times exhibited by both emitters as well as the different profiles in spectra and coherence functions indicate that the immediate environment of the emitters has a varying impact on their photonic properties. In order to emphasize the effect of the environment on photonic coherence, similar measurements performed on a chromium centre implanted directly in bulk diamond with lower density of impurities is shown in both panels as filled blue circles. This centre shows a ZPL at 790 nm. A coherence time longer than 700 ps is extracted for a Gaussian  coherence function (solid red curve) - an order of magnitude longer than the coherence time measured for nanodiamonds.\\
\newline

\begin{figure*}[tbp]
\centering
  \includegraphics[width=0.90\textwidth]{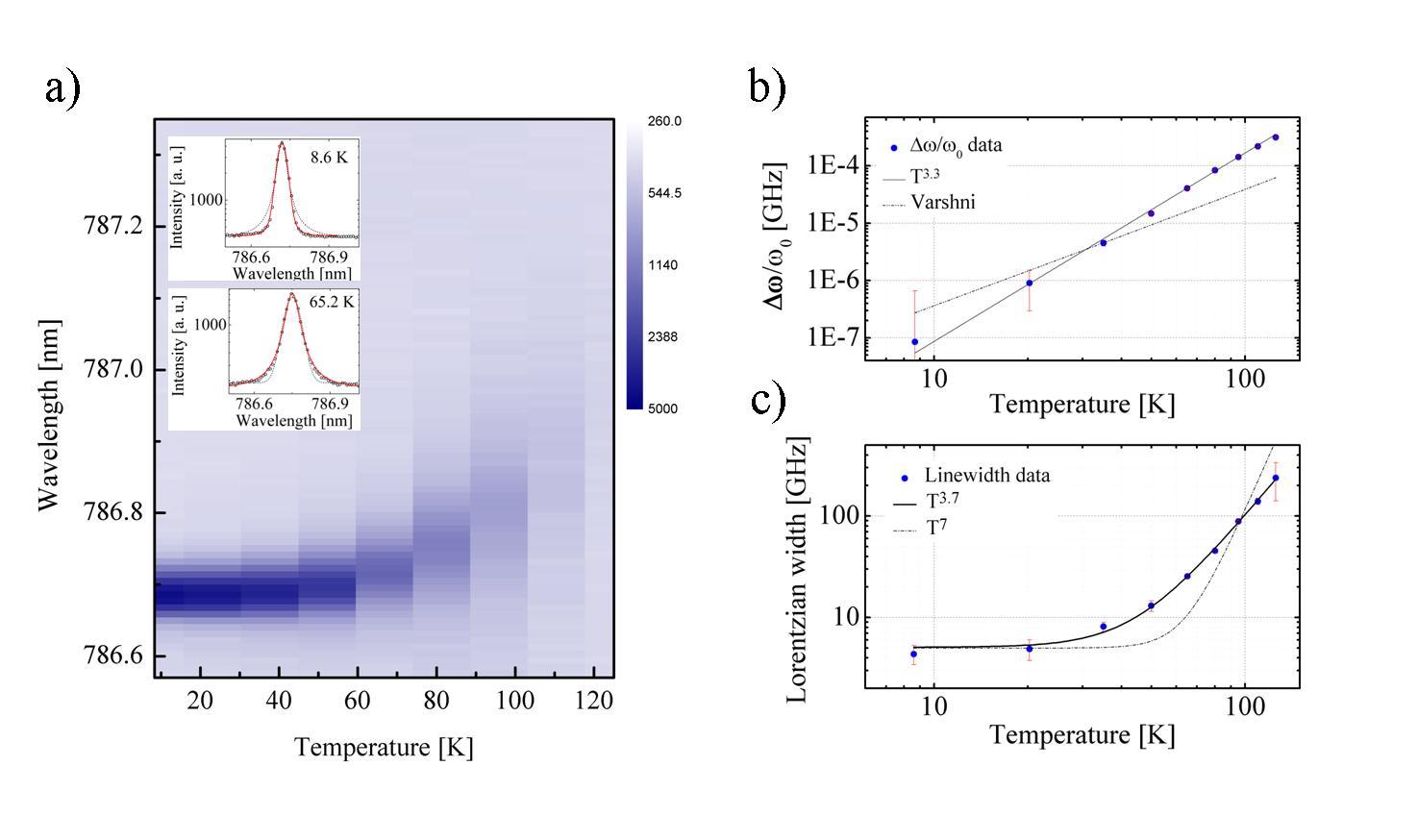}\\
  \caption{(Color online) (a) Temperature dependence of an emission line associated with a chromium impurity for lattice temperatures between 8.6 K and 125.5 K, with temperature increased in steps of 15 K. The insets are spectra of the emitter plotted on a linear-log scale at 8.6 K and 65.2 K, respectively,  and demonstrate the transition from a predominantly Gaussian to a predominantly Lorentzian lineshape. The red curves are Voigt fits and the black dotted curves are a Lorentzian curve at 8.6 K and a Gaussian curve at 65.2 K shown for comparison. (b) The shift of centre frequency follows a $T^{3.3}$ power law (black solid curve). Varshni temperature dependence is shown for comparison (black dotted curve). (c) Transition linewidth as a function of temperature (blue circles).  The linewidth follows a $T^{3.7}$ power law (black solid fit), and a $T^{7}$ dependence is shown for comparison.}
  \label{fig:fig3}
\end{figure*}

The degree of optical dephasing can be revealed in first-order coherence as well as spectral measurements. If the line broadening is small compared to the natural linewidth, the former is more accurate, whereas for strong dephasing, the latter  provides a better means of determining emission linewidths. Figure 3(a) displays the characteristic broadening and red shift of the emission spectrum of another chromium centre (labeled centre C) as a function of lattice temperature between 9 K and 126 K. An analysis of the spectrum reveals that the Gaussian contribution remains constant for all temperatures, while the Lorentzian contribution is temperature dependent. This leads to a switch over from a predominantly Gaussian-like lineshape at low temperatures to a Lorentzian lineshape at higher temperatures for the 10 emitters investigated. This can be seen in the linear-log plots of the lineshape at two different temperatures displayed in the two insets of Fig. 3(a).  The diamond bandgap itself will be modified as a function of temperature, following the Varshni law \cite{Varshni,Varshni_footnote}, and shallow centres will inherit this behaviour. Figure 3(b) displays the fractional change of the transition frequency ($\Delta\omega / \omega$) as a function of temperature, following a power law of the form $\Delta\omega / \omega \propto T^{\alpha}$  with $\alpha$=3.3. This frequency shift is distinctly stronger than the Varshni temperature dependence of  bandgap energy for diamond, as is evident in Fig. \ref{fig:fig3}(b) (black dotted curve). This reveals that all optically relevant states are  sufficiently far in energy from the band edges, as expected for deep defects in diamond. Fig. 3(c) shows the temperature dependence of the linewidth $\gamma$ of the Lorentzian contribution to the emission spectrum of centre C, following a power law of the form $\gamma \propto T^{\beta}$, with $\beta$=3.7. These measurements were repeated for 10 separate centres, and the average values of $\alpha$ and $\beta$ are $3.5\pm0.3$ and $3.4\pm0.3$, respectively. In the low temperature regime data suggests a temperature-independent dephasing mechanism is present for this centre limiting the linewidth to 4.1 GHz. This residual linewidth varies from centre to centre. We note however that a value of 4.1 GHz is within the resolution of the spectrometer ($\sim$ 8 GHz).

A strong candidate for such temperature-dependent broadening mechanism is interaction with lattice phonons. The simplest model for electron-phonon coupling considers quasi-elastic scattering of phonons with a Debye density of states \cite{Maradudin} from a non-degenerate transition of a centre. This model predicts a $T^7$-dependence for dephasing and a $T^4$-dependence for the fractional change of the transition frequency. The strong deviation of the $\beta$ values for chromium centres from the prediction by this model suggests the presence of an additional mechanism for temperatures below the Debye temperature $T_{\textrm{D} }$.  An example of this is already seen in the NV linewidth, which exhibits a $T^5$ behaviour in ultrapure diamond due to Jahn-Teller effect in the excited states \cite{Fu}. Further, experimentally determined values of $\alpha = 3.1$ and $\beta = 3$ have been reported for the N3 colour centre related to nitrogen aggregates in diamond \cite{Halperin}. The particular temperature dependence was attributed to a pronounced deviation of the actual phonon density of states in diamond from the Debye model \cite{Smith}, which was confirmed experimentally for the N3 colour centre using the vibrational structure in the N3 absorption spectrum. A numerical evaluation of linewidth and transition frequency shift using the modified phonon density of states in bulk diamond as a function of temperature according to Ref. \cite{Maradudin} results in $a = 3.1$ and $b = 3.7$, which is in reasonable agreement with the values reported here. This model is plausible for chromium centres as well, although an experimental mapping of the phonon density of states is not feasible due to the small Huang-Rhys parameter \cite{Huang} of these centres, which is less than 0.05 \cite{Siyushev}. That said, the variation of $\alpha$ and $\beta$ from centre to centre suggests that a mechanism governed more by local rather than global properties of the material may be playing a central role in the photon dephasing.
\newline

An alternative model takes into account explicitly the impurity-rich nature of a material and incorporates the effect of phonons to the coupling  of the centres (e.g. Coulombic and spin-dipolar) to nearby lattice impurities and other centres \cite{Hizhnyakov}. Consequently, a time-varying potential at the location of a centre leads to a $T^3$ dependence of the dephasing rate, valid for the temperature range $T_0 < T < T_D$, with $T_D$ the Debye temperature and $T_0 = c^{3/8}T_D$. The coefficient $c$ is the impurity concentration in the host crystal. Applying this general principle to the chromium emitters, an impurity concentration as high as 1 part-per-million  gives $T_0 \sim 10$ K and $T_D = 1850$ K for diamond, so the temperature regime over which the chromium centres are investigated lies within this range. The presence of  charges in the immediate environment of the emitter is already evident from the slow-wandering type decay of coherence exhibited by some of the centres (Gaussian coherence profile in Fig \ref{fig:fig2} (a)). This confirms the plausibility of such a dephasing mechanism. Further investigation of the temperature dependent dephasing as a function of concentration $c$ of defects is necessary for an unambiguous identification of the dephasing type.
\newline

\begin{figure*}[t]
\centering
  \includegraphics[width=0.95\textwidth]{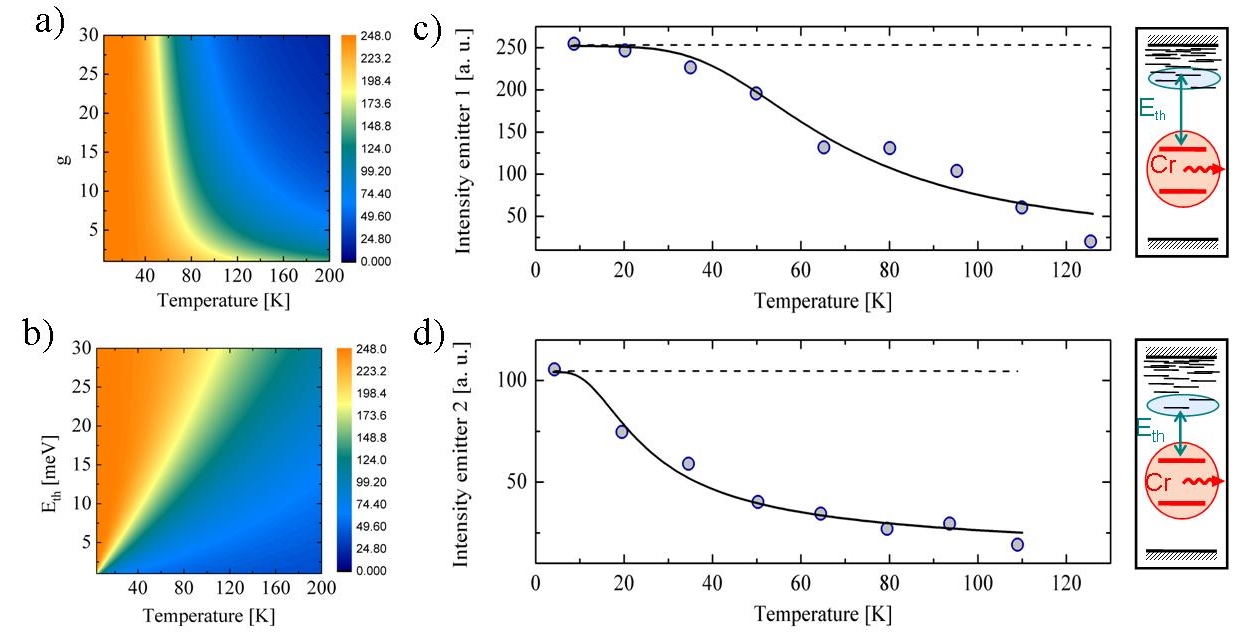}\\
  \caption{(Color online) (a) and (b) ZPL intensity as a function of temperature for varying fitting parameters $g$ and $E_{\textrm{th}}$, respectively. (c) and (d) Intensity of ZPL emission as a function of temperature for chromium emitters C and D, respectively. The side panels show the thermal coupling of the excited state to a reservoir of higher-lying states in the energy bandgap. We note that the centre in panel (c) is the same as that in Fig. 3.}
\end{figure*}

Coupling of a colour centre to a vibronic mode also leads to reduced ZPL intensity with increasing temperature
\cite{Huang, Davies, Warwick}:
\begin{equation}
\begin{split}
I_0(T)=  &I_0(0) \exp{\left[-S \coth\left(\frac{\hbar\omega}{2k_{\textrm{B}}T}\right)\right]}\\
&\times J_0\left[S \, \textrm{csch}\left(\frac{\hbar\omega}{2k_BT}\right)\right]
\label{eq:zplintensity}
\end{split}
\end{equation}

\noindent where $I_0$ is the fraction of emission intensity decaying into the ZPL, $S$ is the Huang-Rhys factor, $\omega$ is the vibronic mode frequency, $\hbar$ is the reduced Planck constant, $k_{\textrm{B}}$  is the Boltzmann constant, and $J_0$ is the zeroeth order Bessel function of the first kind. For chromium centres $S < 0.05$ \cite{Siyushev} and $\hbar\omega$ is on the order of about 10 m\textit{e}V, similar to other diamond centres \cite{Zaitsev, Igor1, Warwick}.
Equation (1) applied to chromium centres yields 95\% of the ZPL intensity to be sustained within the temperature range we study here. In strong contrast, Fig. 3(a) displays the significantly more rapid reduction of the ZPL intensity observed as a function of temperature. Such a strong reduction in intensity is uncharacteristic for $S<0.05$ and suggests an additional mechanism taking part.
\newline

The quantum efficiency, $\eta$, of a transition is determined by the presence of nonradiative channels connecting the excited and the ground states of a centre and is defined as $\eta=P_r/(P_r+P_{nr})$ where $P_r$ ($P_{nr}$) is the probability for a radiative (nonradiative) transition. Nonradiative channels contributing to $P_{nr}$ can become available as a function of temperature via thermal excitation from the excited state of the centre to higher energy nonradiative states. These states can either belong to the centre, or they could be formed by the impurity density in the vicinity of the emitters.  Assuming that the selection rules do not affect the thermalization process, the quantum efficiency becomes temperature dependent:
\begin{equation}
\eta(T)=\frac{1}{1+ge^{-\frac{E_{\textrm{th}}}{k_{\textrm{B}}T}}}.
\label{eq:quantumyield}
\end{equation}

\noindent The nonradiative decay process is assumed here to have a barrier with activation energy $E_{\textrm{th}}$, and the constant $g=g_{\textrm{th}}/g_{\textrm{ex}}$ is the ratio of degeneracies between the thermally activated level and the excited state \cite{Warwick}. The presence of a thermal excitation mechanism to additional states is further evidenced by the bunching behaviour in intensity autocorrelation measurements \cite{Igor1}, which appears predominantly for room-temperature measurements and is suppressed at low temperatures. Including this process, the overall intensity change of the ZPL as a function of temperature is given by
\cite{Warwick, Feng, Collins}:
\begin{equation}
\begin{split}
I_0(T)=&\left(\frac{I(0)}{1+ge^{\left(-\frac{E_{\textrm{th}}}{k_{\textrm{B}}T}\right)}}\right)\exp{\left[-S \coth\left({\frac{\hbar\omega}{2k_{\textrm{B}}T}}\right)\right]}\\&\times J_0\left[S \, \textrm{csch}\left(\frac{\hbar\omega}{2k_{\textrm{B}}T}\right)\right].
\end{split}
\end{equation}
Fig. 4 (a) presents the calculated intensity of the ZPL according to equation 3 as a function of the number degeneracy of the state and the temperature (for a fixed value of the energy separation, $E_{th}$ = 19.6). Similarly, Fig. 4 (b) presents the intensity as a function of $E_{th}$ and temperature for a fixed value of the degeneracy, $g = 19.0$. Clearly, the intensity is sustained until the temperature is high enough to allow thermalization to be significant and the number of degeneracy determines how rapidly the intensity is reduced beyond this temperature. Figures 4c and 4d present experimental results (blue circles) for two separate chromium centres superimposed on the Eq. 3 (solid black curves) using $E_{\textrm{th}}$ and $g$ as fitting parameters. The corresponding values for best fits are given in Table 1 along with those reported for the neutral silicon-vacancy centre \cite{Warwick} for comparison. The values for both $g$ and $E_{\textrm{th}}$ for centre C displayed in Fig. 4(c) are higher than those for centre D displayed in Fig. 4(d). The difference in energy separation and  degeneracy of the thermally occupied states is not surprising given that the chromium emitters are known to be in an environment comprising a high density of impurity atoms such as additional nitrogen, oxygen and silicon \cite{Siyushev, Igor1}. While the model assumes thermal coupling to only one additional state, we interpret from the high values of $g$ and $E_{\textrm{th}}$ that the chromium emitters couple to an ensemble of states with an \textit{effective} energy barrier $E_{\textrm{th}}$ and an overall degeneracy ratio $g$. Intuitively,  the number of states or the degeneracy in the ensemble is expected to increase with larger energy barriers $E_{\textrm{th}}$, since the number of allowed energy states introduced by impurities around the chromium emitters should be higher closer to the valence band edge. This situation is depicted in panels (c) and (d) of Fig. 4. In principle, this model can be extended to include a quasi-continuum band with density of states $g(E)$ formed by the impurities at a mean energy barrier $E_{\textrm{th}}$. Straightforwardly,  $P_{nr}$  in Eq. (2) then becomes  $\int{g(E)\exp(-\frac{E}{k_{\textrm{B}}T})dE}$. We note that using a Gaussian density of states of finite width around a mean energy barrier yields the same functional dependence, however independent measurements would be necessary to reveal the location of the energy barrier per emitter, supported by a density-functional theory calculations of the delocalized electronic states due to the impurities in diamond. The temperature-dependent model we use results in the values of 11\% and 23\% for the room temperature quantum efficiency for centre C and D, respectively. These efficiencies are consistent with previous reports on chromium centres at room temperature \cite{Castelletto} and indicates that the strong suppression of thermal excitation  allows quantum efficiencies at low temperatures, e.g. 4 K, of order unity.
\newline

\begin{table}[ht]
\centering 
\begin{tabular}{c c c c c} 

  & $g$& $E_{\textrm{th}}$ & $\hbar \omega$ & $S$\\ 
\hline 
Centre C	&19.0 $\pm$8	&19.6 $\pm$3.5 meV&	28.9 meV&	0.05 \\
Centre D	&4.3 $\pm$2	&4.4 $\pm$1.7 meV&	28.9 meV&	0.05 \\ 
SiV$^0$	&1-3&5 meV&	28 meV&	1.5 \\ [1ex] 
\hline 
\end{tabular}
\label{table:parameters} 
\caption{Fit parameters for the two chromium centres shown in figure and SiV centres for comparison.} 
\end{table}

In summary, the temperature dependence of photon emission from single chromium centres in diamond revealed the interplay between slow and fast dephasing mechanisms and the nature of coupling to the impurities in the vicinity of the centres.  Due to these coupling mechanisms, the photon coherence times remain about 3.5 times in bulk diamond and 25 times in microcrystals below the theoretical value for transform-limit photon emission. Nonradiative losses due to thermal excitation to additional states are dominant at elevated temperatures, but the quantum efficiency of chromium centres approaches unity at low temperature operation. Deterministic implantation of these centres in ultrapure diamond, or their controlled incoporation in high purity CVD microdiamonds, is therefore needed to make these systems available as good-quality single photon sources.
\newline

{\bf Acknowledgments}

We gratefully acknowledge financial support by the University of Cambridge and the European Research Council (FP7/2007-2013)/ERC Grant agreement no. 209636. Z. Wang and Y. Xin acknowledge the USTC-Cambridge exchange programme for future physicists. We thank C.-Y. Lu for technical assistance, E. Neu for helpful discussions, and D. McCutcheon and A. Nazir for theoretical support.
\newline


\begin{thebibliography}{99}
\bibitem{footnote1} Radiative lifetime broadened transitions for emitters in diamond have so far been observed for the NV centre in ultrapure diamond only \cite{Tamarat}.
\bibitem{Varshni_footnote} The Varshni temperature dependence \cite{Varshni} is of the form $E_{g}(T)=E_{g}(0)-aT^{2}/(T+b)$, where $E_g(0)$=5.4125 eV is the diamond bandgap energy at zero temperature, $a=-1.979 \cdot 10^{-4}$ and $b=-1437$ for bulk diamond.
\end{thebibliography}
\end{document}